\documentclass[12pt]{article}
\usepackage{epsf}
\def\beq{\begin{equation}}
\def\eeq{\end{equation}}

\def\ap#1#2#3 {Ann. Phys. (NY) {\bf#1} (19#2) #3}
\def\apj#1#2#3 {Astrophys. J. {\bf#1} (19#2) #3}
\def\apjl#1#2#3 {Astrophys. J. Lett. {\bf#1} (19#2) #3}
\def\app#1#2#3 {Acta. Phys. Pol. {\bf#1} (19#2) #3}
\def\ar#1#2#3 {Ann. Rev. Nucl. Part. Sci. {\bf#1} (19#2) #3}
\def\cpc#1#2#3 {Computer Phys. Comm. {\bf#1} (19#2) #3}
\def\err#1#2#3 {{\it Erratum} {\bf#1} (19#2) #3}
\def\ib#1#2#3 {{\it ibid.} {\bf#1} (19#2) #3}
\def\jmp#1#2#3 {J. Math. Phys. {\bf#1} (19#2) #3}
\def\ijmp#1#2#3 {Int. J. Mod. Phys. {\bf#1} (19#2) #3}
\def\jetp#1#2#3 {JETP Lett. {\bf#1} (19#2) #3}
\def\jpg#1#2#3 {J. Phys. G. {\bf#1} (19#2) #3}
\def\mpl#1#2#3 {Mod. Phys. Lett. {\bf#1} (19#2) #3}
\def\nat#1#2#3 {Nature (London) {\bf#1} (19#2) #3}
\def\nc#1#2#3 {Nuovo Cim. {\bf#1} (19#2) #3}
\def\nim#1#2#3 {Nucl. Instr. Meth. {\bf#1} (19#2) #3}
\def\np#1#2#3 {Nucl. Phys. {\bf#1} (19#2) #3}
\def\pcps#1#2#3 {Proc. Cam. Phil. Soc. {\bf#1} (#2) #3}
\def\pl#1#2#3 {Phys. Lett. {\bf#1} (19#2) #3}
\def\prep#1#2#3 {Phys. Rep. {\bf#1} (19#2) #3}
\def\prev#1#2#3 {Phys. Rev. {\bf#1} (19#2) #3}
\def\prl#1#2#3 {Phys. Rev. Lett. {\bf#1} (19#2) #3}
\def\prs#1#2#3 {Proc. Roy. Soc. {\bf#1} (19#2) #3}
\def\ptp#1#2#3 {Prog. Th. Phys. {\bf#1} (19#2) #3}
\def\ps#1#2#3 {Physica Scripta {\bf#1} (19#2) #3}
\def\rmp#1#2#3 {Rev. Mod. Phys. {\bf#1} (19#2) #3}
\def\rpp#1#2#3 {Rep. Prog. Phys. {\bf#1} (19#2) #3}
\def\sjnp#1#2#3 {Sov. J. Nucl. Phys. {\bf#1} (19#2) #3}
\def\spj#1#2#3 {Sov. Phys. JETP {\bf#1} (19#2) #3}
\def\spu#1#2#3 {Sov. Phys. Usp. {\bf#1} (19#2) #3}
\def\zp#1#2#3 {Zeit. Phys. {\bf#1} (19#2) #3}

\begin{document}
\begin{titlepage}
\begin{center}
{\Large \bf Theoretical Physics Institute \\
University of Minnesota \\}  \end{center}
\vspace{0.2in}
\begin{flushright}
TPI-MINN-98/29-T \\
UMN-TH-1734-98 \\
December 1998 \\
\end{flushright}
\vspace{0.3in}
\begin{center}
{\Large \bf  On intersection of domain walls in a supersymmetric model
\\}

\vspace{0.2in}
{\bf S.V. Troitsky \\}
Institute for Nuclear Research of the Russian Academy of Sciences,\\
60th October Anniversary Prospect 7a, Moscow 117312 \\[0.1in]
and \\[0.1in]
{\bf M.B. Voloshin  \\ }
Theoretical Physics Institute, University of Minnesota, Minneapolis,
MN
55455 \\ and \\
Institute of Theoretical and Experimental Physics, Moscow, 117259
\\[0.2in]
\end{center}

\begin{abstract}

We consider a classical field configuration, corresponding to
intersection of two domain walls in a supersymmetric model, where the
field profile for two parallel walls at a finite separation is known
explicitly. An approximation to the solution for intersecting walls is
constructed for a small angle at the intersection. We find a finite
effective length of the intersection region and also an
energy, associated with the intersection.
\end{abstract}
\end{titlepage}

\section{Introduction}
The problem of domain walls has a long history both in its general field
aspect \cite{lw} and in a possible relevance to cosmology \cite{zko}.
Recently the interest to properties of domain walls has been given a new
boost in supersymmetric models, which naturally possess several
degenerate vacua, thus allowing for a multitude \cite{mv} of domain wall
type configurations interpolating between those vacua. Furthermore it
has been realised \cite{ds1, ds2} that at least some of these walls have
rather distinct properties under supersymmetry and in fact their field
profiles satisfy first order differential equations in analogy with the
Bogomol'nyi - Prasad - Sommerfeld (BPS) equations \cite{bps}. For this
reason such configurations are called ``BPS-saturated walls", or ``BPS
walls". It was also noticed \cite{shifman} that, quite typically, there
may exist a continuous set of the BPS walls, all degenerate in energy,
interpolating between the same pair of vacua. A further analysis
\cite{sv} of such set in a specific model revealed that the solutions in
the set can be interpreted as two elementary BPS walls parallel to
each other at a finite separation, the distance between the walls being
the continuous parameter labeling the solutions. Since all the
configurations in the set have the same energy, equal to the sum of the
energies of the elementary walls, one encounters here a remarkable
situation where there is no `potential' interaction between the
elementary walls. (This property is protected by supersymmetry and
holds in all orders of perturbation theory.)

\begin{figure}[ht]
  \begin{center}
    \leavevmode
    \epsfbox{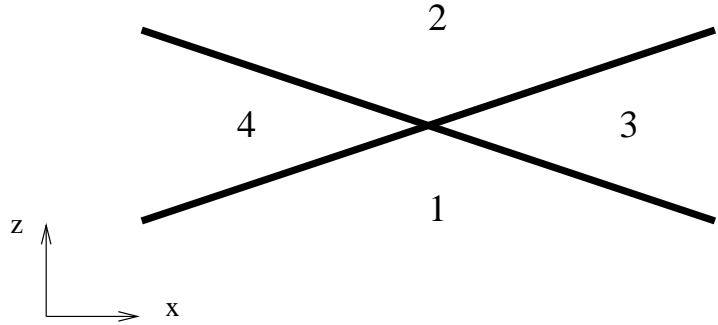}
    \caption{Quadruple vertex for four different vacuum domains}
  \end{center}
\label{fig:cross1}
\end{figure}

The existence of multiple vacua in supersymmetric models naturally
invites a consideration of more complicated field configurations, than
just two vacua separated by a domain wall, namely those with co-existing
multiple domains of different degenerate vacua \cite{mv}. This leads to
the problem of intersecting domain walls. It should be noted, that by
far not every conceivable configuration of domains is stable, e.g. any
intersection of domain walls in a one-field theory is unstable, and the
conditions for stability of the intersections of the walls in
multiple-field theories\cite{mv} generally allow only a well defined set
of intersection angles, depending on relation between the energy
densities of the intersecting walls. As is discussed further in this
paper, for the elementary BPS walls, considered in \cite{sv}, i.e.
non-interacting in the parallel configuration (zero intersection angle
$\beta$), at least a finite range of values for $\beta$, including
$\beta=0$, is allowed by the stability conditions for a quadruple
intersection, shown in Fig.\ref{fig:cross1}. When viewed as an
intersection of world
surfaces of the walls in the space-time, rather than as a static spatial
configuration, the intersection describes scattering of moving walls,
and in the case of a collision of spatially parallel walls (or kinks in
a (1+1) dimensional model) the angle $\beta$ translates into $\beta=v/c$
with $v$ being the relative velocity of the walls, for small $\beta$,
i.e. for non-relativistic collisions. The purpose of the present paper
is a more detailed study of the field profile for the intersection
configuration of such type. We consider here the same supersymmetric
model as in Ref.\cite{sv} and use the explicit solution for parallel
walls, found there, to construct an approximation to the classical field
profile for intersecting walls, which is valid to the first order in
$\beta$ for small $\beta$. This approximation gives the energy of the
static field configuration up to $\beta^2$ inclusive, while for the case
of collision of the walls the corresponding result is obviously the
action for the collision ``trajectory".

Our main findings are as follows. There exists a well defined gap
$\Delta x$ between the vertices of pairwise `meet' of the walls, so that
a more detailed picture of the intersection looks as shown in actual
detail in Fig.\ref{fig:cross3} further in the paper. In the collision
kinematics this
gap is in time and corresponds to a finite time delay in the scattering
process. This classical quantity clearly can also be translated into a
phase shift in a quantum mechanical description of the scattering. The
gap depends on the intersection angle, and is given by
\beq
\Delta x= {1 \over \beta} \, {\ell_0 \over m}~,
\label{gap}
\eeq
where $m$ is a mass parameter for the masses of quanta in the model, and
$\ell_0$ is a dimensionless constant, depending on  the ratio of the
coupling constants in the model. We also find that there is a finite
angle-dependent energy (in the static configuration) associated with the
intersection:
\beq
\varepsilon=\beta  \, \mu \, {\ell_0 \over m}~,
\label{ener}
\eeq
where $\mu$ is the energy density of each of the elementary walls.
Naturally, in a (3+1) dimensional theory $\varepsilon$ is in fact the
energy per unit length of the intersection, while in a (2+1) dimensional
case it is an energy localized at the intersection. In a (1+1)
dimensional model only the collision kinematics (for two elementary
kinks) is possible, thus $\varepsilon$ has the meaning of a finite
action (phase shift) associated with the collision. We believe that the
equations (\ref{gap}) and (\ref{ener}) are quite general at small
$\beta$, and the only model dependence is encoded in the dimensionless
quantity $\ell_0$.

The rest of the paper is organized as follows. In Section 2 we present
the supersymmetric model under consideration and describe the solutions
for the BPS walls. In Section 3 we consider intersecting domain walls
within an {\it Ansatz} allowing us to construct the field profiles in
the limit of small intersection angle, and we find the characteristics
of the intersection. In Section 4 we discuss applicability of the
findings of this paper when some of the restrictions, assumed here, are
relaxed.

\section{Domain walls in a SUSY model}
\begin{figure}[ht]
  \begin{center}
    \leavevmode
    \epsfbox{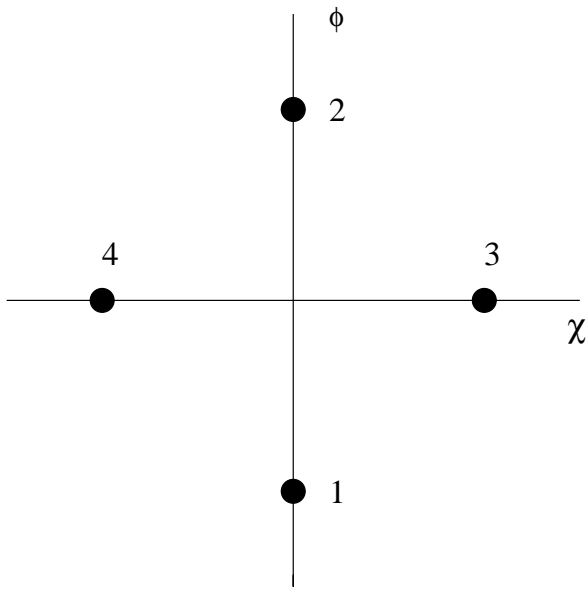}
    \caption{The diagram showing locations of the four vacua and their
labeling in the SUSY model with the superpotential of eq.(\ref{spot}).}
  \end{center}
\label{fig:cross2}
\end{figure}

\begin{figure}[ht]
  \begin{center}
    \leavevmode
    \epsfbox{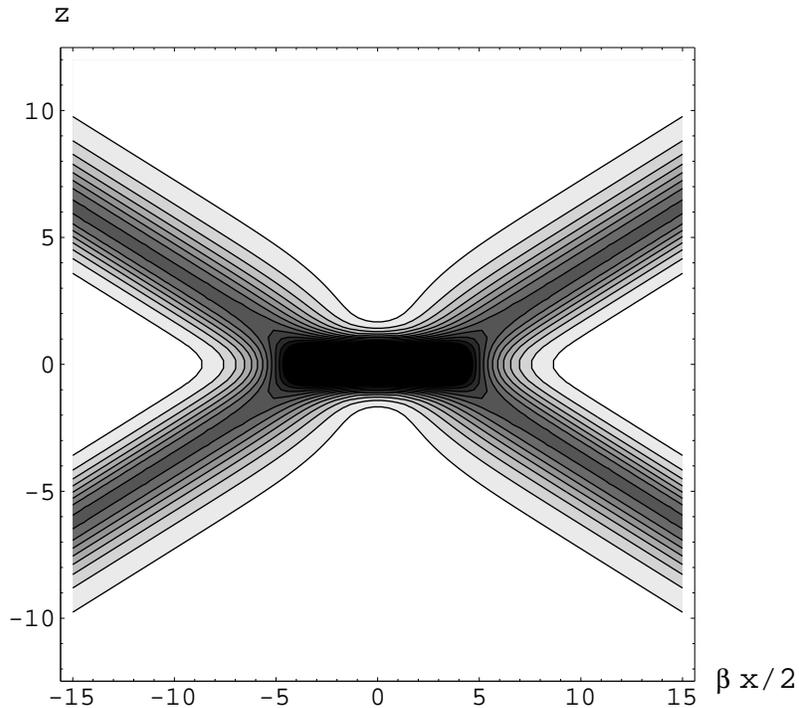}
    \caption{Contour plot of the energy density at the intersection of
the domain walls as described within our {\it Ansatz} in terms of the
scaling variable $\beta \, x/2$. Darker shading corresponds to larger
energy density.}
   \end{center}
\label{fig:cross3}
\end{figure}

The specific SUSY model under consideration is that of two chiral
superfields $\Phi$ and $X$ with the superpotential
\beq
W(\Phi, \, X)= {m^2 \over \lambda} \, \Phi - {1 \over 3} \, \lambda
\,
\Phi^3 - \alpha \, \Phi \, X^2 \, .
\label{spot}
\eeq
Here $m$ is a mass parameter and $\lambda$ and $\alpha$ are coupling
constants. The phases of the fields and of the $W$ are assumed to be
adjusted in such a way that all the parameters are real and positive,
and the lowest components of the superfields $\Phi$ and $X$ are denoted
correspondingly as $\phi$ and $\chi$ throughout this paper. The model
has the $Z_2 \times Z_2$ symmetry under independent flip of the sign of
either of the fields: $\Phi \to -\Phi$, $X \to -X$. The vacuum states in
this model are found as stationary points of the superpotential function
(\ref{spot}): $\partial W(\phi, \, \chi)/\partial \phi =0$ and $\partial
W(\phi, \, \chi)/\partial \chi =0$, and are located at $\phi= \pm
m/\lambda$, $\chi=0$ (labeled here as the vacua 1 and 2), and at
$\phi=0$, $\chi=\pm m /\sqrt{\lambda \, \alpha}$ (the vacua 3 and 4).
The locations and the labeling of the vacuum states are shown in
Fig.\ref{fig:cross2}.
Throughout this paper the fermionic superpartners of the bosons are
irrelevant and also only real components of the fields appear in the
considered configurations. Therefore for what follows it is appropriate
to write the expression for the part of the Lagrangian describing the
real parts of the scalar fields:
\beq
L=\left ( \partial \phi \right )^2 +  \left ( \partial \chi \right )^2 -
\left ( {m^2 \over \lambda} - \lambda \, \phi^2 - \alpha \, \chi^2
\right )^2 - 4 \alpha^2 \, \phi^2 \, \chi^2~.
\label{lagr}
\eeq

As discussed in detail in Ref.\cite{sv}, in this model exists a
continuous set of BPS walls interpolating between the vacua 1 and 2, all
having the energy density $\mu_{12}= {8 \over 3} \, {m^3 \over
\lambda^2}$. These configurations with positive (negative) $\chi$ can be
interpreted as parallel elementary BPS walls, i.e. connecting the
vacua 1 and 3 (1 and 4) and the vacua 3 and 2 (4 and 2) located at a
finite distance from each other. The energy $\mu$ of each of the
elementary walls is $\mu=\mu_{12}/2$. The only non-BPS domain wall in
this model is the one connecting the vacua 3 and 4, and its energy is
given by $\sqrt{\lambda/\alpha} \, \mu_{12}$. The first-order equations
for the BPS walls are conveniently written in terms of dimensionless
field variables $f$ and $h$, defined as
$$
\phi= {m \over \lambda} \, f~, ~~~~~~ \chi= {m \over \sqrt{\lambda \,
\alpha}} \, h~.
$$
For a domain wall perpendicular to the $z$ axis the BPS equations read
as\cite{sv}
\begin{eqnarray}
{{\rm d}f \over {\rm d}z}&=&1-f^2-h^2 \nonumber \\
{{\rm d}h \over {\rm d}z}&=&-{2 \over\rho} \, f \, h~~.
\label{bps}
\end{eqnarray}
Here the notation is used $\rho=\lambda / \alpha$ and the mass parameter
$m$ is set to one.

Although the equations (\ref{bps}) are solved\cite{sv} in quadratures
for arbitrary $\rho$, it is only for $\rho=4$ that the non-trivial
solution can be written in terms of elementary functions. It is this
explicitly solvable case of $\rho=4$ that we consider for the most part
in this paper for the sake of presenting closed expressions wherever
possible. The set of solutions to the equations (\ref{bps}) in this case
reads as\cite{sv}
\beq
f(z)={a \, (e^{2z}-1) \over a+2 e^z + a \, e^{2z}}~, ~~~~h^2 (z)= {2 \,
e^z \over a+2 \,e^z + a \, e^{2z}}~,
\label{sol}
\eeq
where $a$ is a continuous parameter, $0 \le a \le \infty$ labeling the
solutions in the set, and the overall translational freedom is fixed
here by centering the configuration at $z=0$ in the sense that $f(z=0) =
0$. The interpretation of these configurations becomes transparent, if
one introduces the notation:
\beq
\cosh s = a^{-1}~.
\label{sa}
\eeq
Then the expressions (\ref{sol}) can be written as
\beq
f(z)={1 \over 2} \, \left ( \tanh {z-s \over 2} + \tanh {z+s \over 2}
\right )~, ~~~~ h^2(z)={1 \over 2} \, \left (1-  \tanh {z-s \over 2} \,
\tanh {z+s \over 2} \right )~.
\label{solm}
\eeq
In this form it is clear that at large $s$ (i.e. at small $a$) the
functions $f$ and $h$ differ from their values in one of the vacua only
near $z= \pm s$, i.e. the configuration splits into the elementary
walls, corresponding to the transition between the vacua $1 \to 3$
($(f,h)=(-1,0) \to (f,h)=(0,1)$) at $z=-s$ and the transition between
the vacua $3 \to 2$ at $z=s$. (For definiteness we refer to the solution
with positive $h(z)$.) Remarkably, the function $f$ is simply a sum of
the profiles for the elementary walls: $f(z)=f_{13}(z+s)+f_{32}(z-s)$
with
\beq
f_{13}(z)={1 \over 2} \, \left ( \tanh {z \over 2} -1 \right )~, ~~~~
f_{32}(z)={1 \over 2} \, \left ( \tanh {z \over 2} +1 \right )~.
\label{elemf}
\eeq
while the function $h(z)$ is not a linear superposition of the profiles
of $h$ for elementary walls:
\beq
h_{13}^2(z)= {1 \over 2} \, \left ( 1+\tanh {z \over 2} \right )~, ~~~~
h_{32}^2(z)= {1 \over 2} \, \left ( 1-\tanh {z \over 2} \right )~,
\label{elemh}
\eeq
but rather for large $s$ the profile of $h(z)$ is exponentially (in $s$)
close to a linear superposition of the elementary wall profiles
separated by the distance $2s$. Thus at large $s$ (small $a$) the
solution (\ref{sol}) describes two far separated elementary walls.

The caveat of parameterizing the solution in terms of $s$ for arbitrary
positive $a$ is that, according to eq.(\ref{sol}), at $a=1$ the
parameter $s$ bifurcates into the complex plane and becomes purely
imaginary, reaching $s=\pm i \, \pi/2$ at $a=0$. (The functions $f(z)$
and $h(z)$ obviously are still real at $0 \le a \le 1$.)\footnote{It can
be noted that the profile of the fields at $a=1$ is given by the special
solution to the BPS equations first found in Ref.\cite{shifman} at any
$\rho$, such that $\rho > 2$.}

\section{Intersecting domain walls}
In order to describe a static intersection of two elementary walls at a
small angle $\beta$, we make the natural {\it Ansatz} that the parameter
$a$ in the solution (\ref{sol}) is a ``slow" function of the coordinate
$x$, i.e. $f \to f(a(x),z)$ and $h \to h(a(x),z)$. We then substitute
this {\it Ansatz} in the expression for the energy of the fields and
find
\beq
E=\mu \, \int {\rm d} x \, \left [ 2  +  F^2(a) \, \left ( {{\rm d} a
\over {\rm d} x} \right )^2 \right ]
\label{enera}
\eeq
with
\beq
F^2(a)={1 \over (a^2-1)^2} \, \left ( {1 \over a^2} - {5 \over 2} + {3
\, a^2 \arctan \sqrt{a^2-1} \over 2 \, \sqrt{a^2-1}} \right )~.
\label{fa}
\eeq
The interpretation of this expression for the energy becomes quite
simple at small $a$, if one writes it in terms of the ($x$ dependent)
parameter $s$ (cf. eq.(\ref{sa})), assuming that $a < 1$:
\beq
E=\mu \, \int {\rm d} x \, \left [ 2 + {1 \over \sinh^2 s} \, \left (
\cosh^2 s - {5 \over 2} + {3 \, s \over 2 \, \sinh s \, \cosh s} \right
) \, \left ( {{\rm d} s \over {\rm d} x} \right )^2 \right ]~.
\label{eners}
\eeq
At large $s$ the weight function for $({\rm d} s/{\rm d} x)^2$ in the
latter expression rapidly reaches one, and for the trajectory $s=\beta
\, x/2$, corresponding to two walls being at large distance and inclined
towards each other at the relative slope\footnote{At small $\beta$ we
make no distinction between the angle and the slope. When discussing
large $\beta$, or higher order terms, we imply that $\beta$ is the
relative slope, so that the full opening angle between the walls is $2
\, \arctan (\beta/2)$.} $\beta$, the energy per unit length of $x$
coincides in order $\beta^2$ with that of two independent walls: $2\,
\mu \, \sqrt{1+(\beta/2)^2}\,$.

In order to find the relation between $a$ and $x$ at arbitrary
separation between the walls within our {\it Ansatz}, one needs to solve
the variational problem for the energy integral, given by
eq.(\ref{enera}).
The solution is quite straightforward: introduce a function $\sigma(x)$,
such that
\beq
{{\rm d} \sigma \over {\rm d}x}= -F(a) \, {{\rm d} a \over {\rm d} x}
\label{siga}
\eeq
(The minus sign here ensures that $\sigma$ is growing when $a$
decreases.) Then the solution of the variational problem for $\sigma$ is
a linear function of $x$:
\beq
\sigma= \beta \, x/2~.
\label{sigx}
\eeq
The overall shift in $x$ and $\sigma$ is chosen so that $\sigma=0$ and
$x=0$ when $a = \infty$ i.e. at the center of the intersection. The
slope of the $\sigma(x)$ is determined by noticing that at $a \to 0$ the
function $F(a)$ behaves as $F(a)=a^{-1}+O(a)$, thus at small $a$ the
slope of $\sigma$ coincides with that of $s(x)$, defined above as
$\beta/2$ at large $s$.

At this point we address the question about the accuracy of our {\it
Ansatz} at small $\beta$, and consider the full second-order
differential equations for the fields $\phi$ and $\chi$ following from
the Lagrangian in eq.(\ref{lagr}). Since at fixed $a$ the profile given
by eqs.(\ref{sol}) satisfies also the second-order equations in the $z$
variable, the mismatch in the full two-dimensional equations is given by
the second derivatives in $x$ only: $\partial^2 f(a(x),z) /\partial x^2$
and $\partial^2 h(a(x),z) /\partial x^2$. One can readily see however
that these quantities are of order $\beta^2$. Indeed, e.g. for the term
with $f(a(x),z)$ one finds using eqs.(\ref{siga}) and (\ref{sigx})
$$
{\partial^2 f(a(x),z) \over \partial x^2}= {\beta^2 \over 4 \, F^2(a)}
\, \left ( {\partial^2 f(a,z) \over \partial a^2}- {F^{'}(a) \over F(a)}
\, {\partial f(a,z) \over \partial a} \right )~.
$$
Thus our {\it Ansatz} correctly approximates the actual solution in the
first order in $\beta$. Once it is established that the correction
$\phi_i^{(2)}$ to the solution within the {\it Ansatz} is $O(\beta^2)$
(with $\phi_i$ generically denoting the fields $\phi$ and $\chi$), it is
clear that the corrections to the energy can start only in the order
$\beta^4$. Indeed, the variation of the energy, minimized within the
{\it Ansatz}, $\delta E / \delta \phi_i$ is of order of the mismatch in
the field equations, i.e. $O(\beta^2)$. Thus the error in the found
energy
$(\delta E / \delta \phi_i) \, \phi_i^{(2)}$ is $O(\beta^4)$.

A remark concerning further details of the discussed solution is due in
relation with the singularity in $a$ at $x=0$. Formally, the described
solution is only specified so far at $x \ge 0$. At negative $x$ one a
priori can choose one of two options: symmetrically reflect the profile
at positive $x$, or also flip the sign of the field $h$ at negative $x$.
The first option would correspond to the domains of the same vacuum
(i.e. the vacuum 3 or the vacuum 4) at small $z$ on both sides of the
intersection, while the second option describes the change from the
vacuum 3 to the vacuum 4 at the intersection. The first configuration
however is unstable (the translational zero mode develops a nodal line),
and only the second type configuration should be chosen.

One can notice that the profile of the fields within our {\it Ansatz}
depends in fact on the scaling variable $\sigma= \beta \, x/2$. This
obviously implies that a `longitudinal' interval of $\delta x$ between
some fixed characteristic values of the fields in the configuration
scales as $\beta^{-1}$. The behavior of the parameter $a$ in the
expression (\ref{sol}) for the field profile as a function of $\sigma$
is determined by the equation
\beq
F(a) \, {{\rm d} a \over {\rm d} \sigma}=-1~.
\label{asig}
\eeq
This equation can be readily solved numerically, and the resulting
picture of the intersection is presented in Figures 3 and 4.

\begin{figure}[ht]
  \begin{center}
    \leavevmode
    \epsfbox{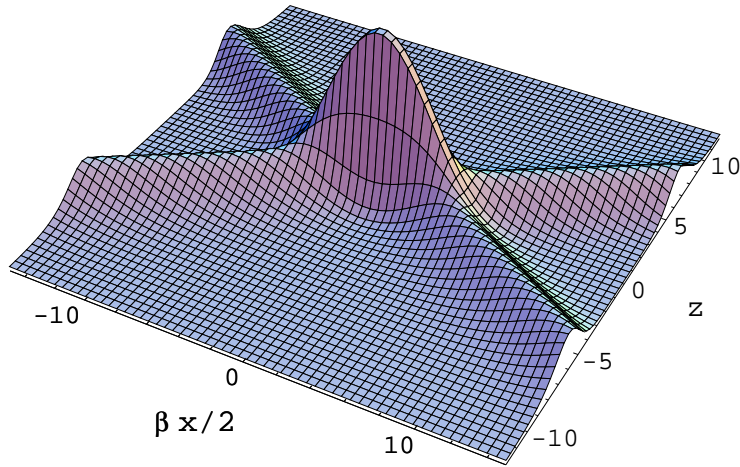}
    \caption{Three dimensional plot of the energy density at the
intersection.}
  \end{center}
\label{fig:cross4}
\end{figure}

It is seen from the figures that the four crests of the energy density
profile do not intersect at one point, rather there is a finite gap
between the pairwise intersections of the energy crests for the
`initial' and the `final' walls. In order to quantify this gap, we find
from eqs.(\ref{asig}) and (\ref{sa}) the relation between $\sigma$ and
$s$ as
\beq
\sigma= \int^\infty_{1 / \cosh s} F(a) \, {\rm d}a
\label{sigmas}
\eeq
and then find numerically that at large $s$ this relation gives
$\sigma=s+\delta $ with $\delta =1.7527\ldots$. Since at large $s$ the
distance between the elementary walls is naturally identified as $2s$,
one concludes that when the intersection of two walls is extrapolated
from large distances it comes short of the actual center of the solution
by the distance $2 \delta/\beta$. The gap $\ell_0$ between the apparent
pairwise collision vertices is then twice this distance. In this way we
obtain the equation (\ref{gap}) with $\ell_0=4 \, \delta = 7.0108\ldots$
in the particular model with $\rho=4$.

In order to find the proper expression for the energy associated with
the intersection, one can consider the problem of intersecting walls as
a boundary value problem in a large box with the size $L$ in the $x$
direction (so that the boundaries are at $x = \pm L/2$), and at the
boundaries the positions of the walls in the $z$ direction are specified
as $s=\pm l/2$, where $s$ is the parameter in the expression
(\ref{solm}) for the profile. (Clearly  the sign of $h(z)$ at $x=+L/2$
should be chosen opposite to that at $x=-L/2$ in order to have an
intersection configuration.) We also assume that the separation $l$
between the walls at the boundaries satisfies $l \ll L$ in order to
correspond to $\beta \ll 1$. For free walls, with no energy associated
with the intersection, the energy of such configuration would be
determined by the total length of the walls in the box:
\beq
E_0=2 \, \mu \, \sqrt{L^2 + l^2} \approx 2 \, \mu \, L + \mu \, L \,
\left ( { l \over L} \right )^2~.
\label{ener0}
\eeq
For the solution within our {\it Ansatz}, however, the energy is given
by
\beq
E=2 \, \mu \, L + \mu \, L \, {\beta^2 \over 4}~,
\label{ener1}
\eeq
where $\beta /2$ is the slope in the solution $\sigma(x)=\beta \, x/2$. 
At $x=L/2$ one can use the asymptotic relation $\sigma = s + \delta$,
and find from the boundary condition for $s$ the relation for $\beta$:
$$
{\beta \over 2}= {l \over L} + {2 \, \delta \over L}~.
$$
Upon substituting this relation in eq.(\ref{ener1}) one finds a finite
difference between the expressions (\ref{ener1}) and (\ref{ener0}) for
the energy: $\varepsilon=E-E_0$, given by our result in eq.(\ref{ener}).

If instead of the spatial coordinate $x$ one uses the time coordinate
$t$, the discussed configuration describes a collision of two parallel
walls (or two kinks in (1+1) dimensional theory). In this case the
parameter of the intersection $\beta$ is identified as the relative
velocity $v$ of the walls, and eq.(\ref{gap}) describes the time delay
in the process of scattering of the walls, while the quantity
$\varepsilon$ in eq.(\ref{ener}) gives the additional action with
respect to the free motion of the walls, and thus is equal to the phase
of the transmission amplitude in a quantum mechanical description of the
scattering.

\section{Discussion}
The description of intersection of elementary domain walls is found here
for small intersection angles and for a particular value of the
parameter $\rho$, $\rho=4$, in a specific SUSY model. At present we can
only speculate how our results are modified beyond these restrictions.
Although the same {\it Ansatz}, as used here, can be applied in the
limit of small $\beta$ at any value of $\rho$, one inevitably has to
resort to numerical analysis because of lack of an explicit simple
expressions for the profiles of the fields, except for the trivial case
$\rho=1$.
In particular, the existence of the gap at the intersection appears to
be related to the existence of the special bifurcation
solution\cite{shifman} for parallel walls. Indeed a non-zero difference
$\delta$ between $\sigma$ and $s$, arising from the integral in
eq.(\ref{sigmas}), comes mostly from the integration over the values of
$a$ past the bifurcation point, i.e. from $a=1$ to $a=\infty$. The
bifurcation solution on the other hand exists in the considered model
only for $\rho \ge 2$, and at $\rho=2$ the solution corresponds to
$\chi(z)=0$, so that the bifurcation point is located at $x=0$ for such
$\rho$. Our preliminary numerical study of the dependence of the size of
the gap on $\rho$ indicates that indeed the gap appears to arise
starting with $\rho=2$ and becomes larger with increasing $\rho$,
whereas at $\rho < 2$ we have found no apparent gap. This conjecture
about such behavior is somewhat supported by the fact that at $\rho=1$
the elementary walls are not interacting and simply ``go through" at the
intersection with no gap whatsoever.

As to the dependence on the intersection parameter $\beta$, our {\it
Ansatz} becomes inapplicable when $\beta$ is not small, and the behavior
of the corresponding configurations is not known. Here we can only
mention that the stability conditions\cite{mv} for the wall
intersections allow a static  quadruple intersection with any $\beta$ if
$\rho >1$. Indeed in this case the energy of the non-BPS wall ($3 \to
4$), $\mu_{34}=2 \, \sqrt{\rho} \, \mu$ is larger than the sum of the
energies of the elementary walls, and thus the quadruple intersection
cannot split into triple ones, since the stability conditions for triple
intersections can not be satisfied. On the contrary, at $\rho <1$ such
splitting is possible with the critical value of $\beta$ defined as
$\beta_c/(2 \, \sqrt{1+(\beta_c/2)^2})= \sqrt{\rho}$.

BPS solitons are known to present effective low-energy degrees of
freedom in some supersymmetric field theories. These effective
(``dual'') theories exploit the fact that elementary solitons do not
interact with each other at zero energies, both classically and quantum
mechanically. This is just the case in the model considered, where the
energy of parallel elementary domain walls does not depend on distance
between them, and this degeneracy is not lifted by quantum corrections.
We demonstrated here by an explicit calculation that some highly
nontrivial interaction between BPS domain walls arises at nonzero
momenta. If this observation holds for general case, to obtain dynamical
information from dual models of that class appears to be an extremely
complicated task.

\section{Acknowledgements}
One of us (SVT) acknowledges warm hospitality of the Theoretical Physics
Institute at the University of Minnesota, where this work was done.
The work of SVT is supported in part by the RFFI grant 96-02-17449a and
in part by the U.S. Civilian Research and Development Foundation for
Independent States of FSU (CRDF) Award No. RP1-187.
The work of MBV is supported in part by DOE under the grant number
DE-FG02-94ER40823.


\begin{thebibliography}{99}

\bibitem{lw}
T.D. Lee and G.C. Wick, \prev{D9}{74}{2291}.
\bibitem{zko}
Ya.B. Zeldovich, I.Yu. Kobzarev, and L.B. Okun, Zh.Eksp.Teor.Fiz. {\bf
67} (1974) 3 (\spj{40}{74}{1}).
\bibitem{mv}
M.B. Voloshin, \prev{D57}{98}{1266}.
\bibitem{ds1}
G. Dvali and M. Shifman, \np{B504}{97}{127}.
\bibitem{ds2}
G. Dvali and M. Shifman, \pl{B396}{97}{64}, \ Err.-ibid. {\bf B407}
(1997) 452.
\bibitem{bps}
E. Bogomol'nyi, \sjnp{24}{76}{449};\\
M.K. Prasad and C.H. Sommerfeld, \prl{35}{76}{760}.
\bibitem{shifman}
M. Shifman, \prev{D57}{98}{1258}.
\bibitem{sv}
M.A. Shifman and M.B. Voloshin, \prev{D57}{98}{2590}.


\end{thebibliography}
\end{document}